\documentclass[doublespacing]{elsart}
\usepackage{graphicx}
\usepackage{subfigure}
\usepackage{epsfig}
\usepackage{amsmath}
\usepackage{amsfonts}   
\usepackage{amssymb}
\journal{...}

\newcommand{\mlabel}[1]{\label{#1}
}

\newcommand{\seq}{\begin{equation}}                 
\newcommand{\eeq}[1]{\label{#1}\end{equation}
    }

\newcommand{\epf}{$ \quad \Box$ \par \vspace{1ex}}

\newtheorem{Theorem}{Theorem}[section]
\newcommand{\sthm}{\begin{Theorem}}         
\newcommand{\ethm}{\end{Theorem}}           

\newtheorem{Corollary}[Theorem]{Corollary}
\newcommand{\scor}{\begin{Corollary}}       
\newcommand{\ecor}{\end{Corollary}}         

\newtheorem{Lemma}[Theorem]{Lemma}
\newcommand{\slm}{\begin{Lemma}}            
\newcommand{\elm}{\end{Lemma}}              

\newtheorem{Proposition}[Theorem]{Proposition}
\newcommand{\spro}{\begin{Proposition}}            
\newcommand{\epro}{\end{Proposition}}              

\newtheorem{Example}[Theorem]{\sc Example}
\newcommand{\sex}{\begin{Example}\rm}        
\newcommand{\eex}{\end{Example}}             

\newtheorem{Remark}[Theorem]{Remark}
\newcommand{\srmark}{\begin{Remark}\rm}        
\newcommand{\ermark}{\end{Remark}}             

\newtheorem{Definition}[Theorem]{Definition}
\newcommand{\strdef}{\begin{Definition}\rm}        
\newcommand{\eeddef}{\end{Definition}}             

\newcommand{\seql}{\begin{eqnarray*}}       
\newcommand{\eeql}{\end{eqnarray*}}

\newcommand{\smlist}[1]{\begin{list}           
                      {(#1{zzcount})}{\usecounter{zzcount}}}
\newcommand{\elist}{\end{list}}

\newcommand{\vect}[1]{\mathbf{#1}}
\providecommand{\abs}[1]{\lvert#1\rvert}
\providecommand{\norm}[1]{\lVert#1\rVert}

\begin{document}
\begin{frontmatter}

\title{Spreading speeds and traveling waves for a model of epidermal wound healing\thanksref{ref1}}
\author{Haiyan Wang}
\address{Division of Mathematical and Natural Sciences\\ Arizona State University \\ Phoenix, AZ 85069-7100, USA}
\thanks[ref1]{Supported by a SRCA grant from the New College of Interdisciplinary Arts and Sciences at Arizona State University}
\ead{wangh@asu.edu}
\begin{abstract}
In this paper, we shall establish the spreading speed and existence of traveling waves for a non-cooperative system arising from epidermal wound healing
and characterize the spreading speed as the slowest speed of a family of non-constant traveling wave solutions. Our results on the spreading speed and
traveling waves can also be applied to a large class of  non-cooperative reaction-diffusion systems.
\end{abstract}

\end{frontmatter}

\pagenumbering{arabic}

\section{Introduction}\label{induc}

In this paper, we study the spreading speeds and traveling wave solutions of  a non-cooperative reaction-diffusion systems arising from wound healing.
Wound healing is complex and remains only partially understood, despite extensive research. Several reaction-diffusion
models have been developed in Sherratt and Murray \cite{Sherratt1990,Sherratt1991},
Dale, Maini, Sherratt \cite{Dale1994} and others to understand the biological process of epidermal wound healing through
mathematical analysis and numerical simulations. We refer to Murray \cite{murray2003} for more detailed discussions and further references.
The models consist of two conservation equations, one for the epithelial cell density per unit area ($u_1(x,t)$)
and one for the concentration of the mitosis-regulating chemical ($u_2(x,t)$). There are two types of the chemicals,
one in which the chemical activates mitosis and the other in which it inhibits it. The following
simplified model was proposed in \cite{Sherratt1990,murray2003} for the activator
\begin{equation}\label{eqoper00}
\begin{split}
\frac{\partial u_1}{ \partial t}&=d_1 \Delta u_1+s(u_2)u_1(2-u_1)-u_1\\
\frac{\partial u_2}{ \partial t}&=d_2\Delta u_2+b(h(u_1)-u_2)\\
\end{split}
\end{equation}
where $b>0, \kappa \in (0, \frac{1}{2})$, $s(u_2)=\kappa u_2+1-\kappa$ is the linearized function which reflects the chemical control of motosis. The chemical production by the function
$h(u_1)=\frac{u_1(1+\zeta^2)}{u_1^2+\zeta^2}, \zeta \in (0,1)$ reflects an appropriate cellular response to injury.
The qualitative form of the solution of (\ref{eqoper00}) in the linear phase is of a wave moving with constant shape and speed. Such a solution is amenable to analysis
if we consider a cone dimensional geometry rather than the two dimensional radially symmetric geometry. Mathematically, we look for a traveling wave
solution of the form $u_1(x,t)=u_1(\xi), u_2(x,t)=u_2(\xi), \xi=x+ct$ where $c$ is the wave speed, positive since here we consider waves moving to
the left.  In Section \ref{example}, we shall establish the existence of traveling waves as well as the results on the speed of propagation to (\ref{eqoper00}).
In addition,
we characterize  the minimum speed as the slowest speed of a family of non-constant traveling wave solutions of (\ref{eqoper00}).

Traveling wave solutions and spreading speeds for reaction-diffusion equations have been studied by a number of researchers. Fisher \cite{Fisher} studied the nonlinear parabolic equation
\begin{equation}\label{eq0}
w_t=w_{xx}+w(1-w).
\end{equation}
for the spatial spread of an advantageous gene in a population and conjectured $c^*$ is the asymptotic speed of propagation of the advantageous gene.
His results show that (\ref{eq0}) has a traveling wave solution of the form $w(x+ct)$ if only if $|c| \geq c^*=2.$ Kolmogorov, Petrowski, and Piscounov
\cite{Kolmogorov} proved the similar results with more general model. Those pioneering work along with the paper by
Aronson and Weinberger \cite{Aronson1975,Aronson1978} confirmed the conjecture of Fisher and established the speeding spreads for nonlinear parabolic equations.
Lui \cite{Lui1989} established the theory of spreading speeds for cooperative recursion systems. In a series of papers,
Weinberger, Lewis and  Li \cite{Weinberger2002,Weinberger2005,Weinberger2002-1,Weinberger2007} studied spreading speeds and traveling waves for more general
cooperative recursion systems, and in particular, for quite general cooperative reaction-diffusion systems by analyzing of
traveling waves and the convergence of initial data to wave solutions. However, mathematical challenges remain because many reaction-diffusion systems are not necessarily cooperative due to  various
biological or physical constrains. Thieme \cite{Thieme1979} showed that asymptotic spreading speed of integral equations with
nonmonotone growth functions can still be obtained by constructing monotone functions. For a related nonmonotone integro-difference equation, Hsu and Zhao \cite{hsu2008},
Li, Lewis and Weinberger \cite{LiLewis2009} extended the theory of spreading speed and established the existence of travel wave solutions. The author and Castillo-Chavez \cite{HwangIntegralDiff}
prove that a class of nonmonotone integro-difference systems have spreading speeds and traveling wave solutions.
Such an extension is largely based on the construction of two monotone operators with appropriate properties and fixed point theorems in Banach spaces.
A similar method was also used in Ma \cite{ma2007} and the author \cite{Hwang2009} to
prove the existence of traveling wave solutions of nonmonotone reaction-diffusion equations. Weinberger, Kawasaki and Shigesada \cite{WeinbergerKS2009} discuss the minimum spreading speeds for a partially-cooperative system describing
the interaction between ungulates and grass. It is cooperative for small population densities but not for large ones.  By employing comparison
methods \cite{WeinbergerKS2009} established the spreading speeds of propagation.  In a recent paper \cite{HwangSystemPDE},
we study traveling waves and spreading speeds of propagation for a class of non-cooperative reaction-diffusion systems and a slightly different model
describing the interaction between ungulates and grass.

In this paper, we shall study the spreading speeds and existence  of traveling waves for the non-cooperative system arising from epidermal wound healing
(\ref{eqoper00}). The minimum speed $c^*$ can be characterized as the slowest speed of a family of non-constant traveling wave solutions. In other words,
we shall show that for $c\geq c^*$ (\ref{eqoper00}) always has a nonconstant traveling solutions of the form $(u_1(x+ct),u_2(x+ct))$ with
$(u_1(-\infty),u_2(-\infty))=0$ but bounded away from zero at $+\infty$,
and there is no such traveling solution when $0 \leq c<c^*.$  Our main results
for the epidermal wound healing model are summarized in Theorem \ref{th33}. The results for general non-cooperative systems (\ref{eq1}) are
included in Theorem \ref{th20}. In Section \ref{example}
we shall verify the assumptions of Theorem \ref{th20} for (\ref{eqoper00}) and apply the general results to (\ref{eqoper00}).

In order to better understanding of the spreading speeds and traveling solutions for the epidermal wound healing model. The general results in this paper
have some significant improvements of those results over \cite{HwangSystemPDE}.
For example, in this paper we make use of the comparison principle from Fife \cite{Fife1979}.
Another form of the comparison principle from \cite{WeinbergerKS2009} was used in \cite{HwangSystemPDE}. As a result, the assumptions (H1-H2) and
the proofs in Section \ref{spreadings}  are somewhat
different from \cite{HwangSystemPDE}. By a suitable modification of the functions, the conditions in this paper seem easier to verify. In addition, Theorem \ref{th30} is
more general than that in \cite{HwangSystemPDE}, for example, $d_1 \geq d_2$ is imposed in \cite{HwangSystemPDE}. To take $d_1 < d_2$ into consideration, some assumptions and
proofs are substantially  modified. In particular, for this epidermal wound healing model, we show that the condition for Theorem \ref{th30}(ii)
can be satisfied if $u_1(x) \not\equiv 0$. Finally, verifications of lower and upper solutions for the equivalent integral equations are
significantly simplified via a result in Ma \cite{ma2001}.
In \cite{Hwang2009} and \cite{HwangSystemPDE}, a more direct but lengthy verification of the lower and upper solutions
are given for scalar and n-dimensional systems respectively. We also omit some standard proofs such as continuity and compactness for the operator
which can be found in previous papers.

\section{Preliminaries}\label{preli}
We begin with some notation. We shall use $R, k, k^{\pm}, f, f^{\pm}, r, u,v$ to denote vectors in $\mathbb{R}^N$ or $N$-vector valued functions
, and $x,y,\xi$ the single variable in $\mathbb{R}$. Let $u=(u_i), v=(v_i) \in \mathbb{R}^N$, we write $u \geq v$ if $u_i \geq v_i $ for all $i$;
and
$u \gg v$ if $u_i > v_i$ for all $i$. We further define for any $r=(r^i)>>0, r \in \mathbb{R}^N$  the ${R}^N$-interval
$$
[0,r]= \{ u: 0 \leq u \leq r, u \in \mathbb{R}^N\}\subseteq \mathbb{R}^N
$$
and
$$
\mathcal{C}_{r}= \{u=(u_i): u_i \in C(\mathbb{R}, \mathbb{R}), 0\leq u_i(x) \leq r_i,  x\in \mathbb{R},\; i=1,...,N\},
$$
where $C(\mathbb{R}, \mathbb{R})$ is the set of all continuous functions from $\mathbb{R}$ to $\mathbb{R}$.

Consider the system of reaction-diffusion equations
\begin{equation}\label{eq1}
u_t=Du_{xx}+f(u), x \in \mathbb{R},\; t\geq 0.
\end{equation}
with
\begin{equation}\label{eq1bc}
u(x,0)=u_0(x),\;\;  x \in \mathbb{R},
\end{equation}
where $u=(u_i)$, $D=\text{diag} (d_1, d_2, ...,d_N), d_i>0, i=1,...,N$
$$f(u)=(f_1(u),f_2(u),...,f_N(u)),$$
$u_0(x)$ is a bounded uniformly continuous function on $\mathbb{R}.$ In this paper, by a solution we mean a twice continuously differentiable
function $u(x,t)$ in $\mathbb{R} \times (0,\infty)$  and continuous in $\mathbb{R} \times [0,\infty)$, and satisfying appropriate equation in
$\mathbb{R} \times (0,\infty)$ and an initial condition.

In order to deal with non-cooperative system, we shall assume that there are additional
two monotone operators $f^{\pm}$, one lies above and another below $f$ with the corresponding equations
 \begin{equation}\label{eq1+}
u_t=Du_{xx}+f^+(u), x \in \mathbb{R},\; t\geq 0.
\end{equation}
\begin{equation}\label{eq1-}
u_t=Du_{xx}+f^-(u), x \in \mathbb{R},\; t\geq 0.
\end{equation}
Such an assumption will enable us to make use of the corresponding results for cooperative systems in
\cite{Lui1989, Weinberger2002-1} to establish spreading speeds for (\ref{eq1}).
\begin{enumerate}
    \item[(H1)] \begin{itemize}
             \item[i.] Let $f, f^{\pm}: \mathbb{R}^N \to \mathbb{R}^N$ be Lipschitz continuous, twice piecewise continuous differentiable function such that
                           \begin{equation*}
              f^{-}(u) \leq f(u) \leq f^{+}(u), u \in \mathbb{R}^N.
              \end{equation*}
              \item[ii.] Let $ 0<<k^{-}=(k^-_i)\leq k=(k_i) \leq k^{+}$ and $f(0)=f(k)=0$ and assume that there is no other positive equilibrium of $f$ between $0$ and $k$ (that is, there is no constant
              $v \neq k$ such that $f(v)=0, 0 << v \leq k$).
              $f^{\pm}(0)=f^{\pm}(k^{\pm})=0$. There is no other positive equilibrium of $f^{\pm}$ between $0$ and $k^{\pm}$.
              \item[iii.] (\ref{eq1+}) and (\ref{eq1-}) are cooperative
              (i.e. $\partial_if^{\pm}_j(u) \geq 0$ for $u \in [0, k^{+}], i\neq j$). $f^{\pm}(u), f(u)$ have the same Jacobian matrix $f'(0)$ at $u=0$.
              \end{itemize}
\end{enumerate}

A traveling wave solution $u$ of (\ref{eq1}) is  a solution of the form $u=u(x+ct), u \in C(\mathbb{R}, \mathbb{R}^N)$ .
Substituting $u(x,t)=u(x+ct)$ into (\ref{eq1}) and letting $\xi=x+ct$, we obtain the wave equation
\begin{equation}\label{eq211}
Du''(\xi)-cu'(\xi)+f(u(\xi))=0, \;\; \xi \in \mathbb{R}.
\end{equation}
Now if we look for a solution of the form
$(u_i)=\big ( e^{\lambda \xi}\eta^i_{\lambda}\big),\lambda>0, \eta_{\lambda}=(\eta^i_{\lambda})>>0$ for the linearization of (\ref{eq211}) at the origin,
we arrive at the following system equation
\begin{equation*}
\text{diag}(d_i \lambda^2 -c \lambda)\eta_{\lambda}+f'(0)\eta_{\lambda}=0
\end{equation*}
which can be rewritten as the following eigenvalue problem
\begin{equation}\label{egenvalue}
\frac{1}{\lambda}A_{\lambda}\vect{\eta_{\lambda}}= c \vect{\eta_{\lambda}},
\end{equation}
where
\begin{equation*}
A_{\lambda}=(a^{i,j}_{\lambda})=\text{diag}(d_i\lambda^2)+f'(0)
\end{equation*}

The matrix $f'(0)$ has nonnegative off diagonal elements. In fact, there is a constant $\alpha$ such that $f'(0)+\alpha I$ has
nonnegative entries, where $I$ is the identity matrix.

By reordering the coordinates, we can assume that $f'(0)$ is in block lower triangular
form, in which all the diagonal blocks are irreducible or $1$ by $1$ zero matrix. A matrix is irreducible if it is not similar
to a lower triangular block matrix with two blocks via a permutation. Let $\rho(A)$ be the spectral radius of $A$. For a matrix $A$
with nonnegative off diagonal elements, from the Perron-Frobenius theorem, we shall call the eigenvalue
$$\Psi(A)=\rho(A+\alpha I)-\alpha$$
of $A$, which has the same eigenvector, the principal eigenvalue of $A$ (see e.g. \cite{Horn,Weinberger2002-1}).
Here $A+\alpha I$ is nonnegative, and $\rho(A+\alpha I)$ is the spectral radius of $A+\alpha I.$

 We make assumptions on $A_{\lambda}$ and also requires $f$ grows less than its linearization along
 the particular function $\nu_{\lambda} e^{-\lambda x }$ \cite{Weinberger2002-1}. Such a condition can be satisfied for many biological systems.

\begin{figure}
\begin{center}
  \includegraphics[width=10cm]{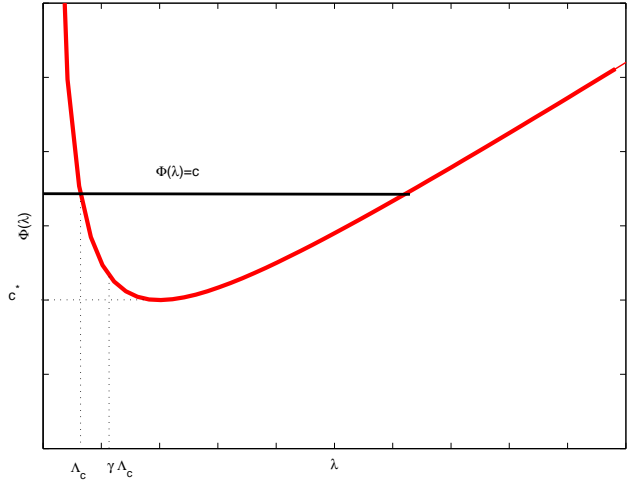}
  \caption{The red curve is $\Phi(\lambda)$. The minimum of $\Phi(\lambda)$ is $c^*$. For $c>c^*$, the left solution of $\Phi(\lambda)=c$ is $\Lambda_c$.} \label{fig4}
\end{center}
\end{figure}

\begin{enumerate}
   \item[(H2)] \begin{itemize}
   \item[i] Assume that $A_{\lambda}$ is in block lower triangular
form and the first diagonal block has the positive principal eigenvalue  $\Psi(A_{\lambda})$, and $\Psi(A_{\lambda})$ is strictly larger than the principal
eigenvalues of all other diagonal blocks for some interval $[0, \Lambda^*]$ of $\lambda$. Assume that for $\lambda \in [0, \Lambda^*]$ (we also assume that $\Lambda_{c^*} \leq \Lambda^* \leq \infty$, see
    Lemma \ref{lmeigen} for $\Lambda_{c^*}$), there is a positive eigenvector $\nu_{\lambda}=(\nu^i_{\lambda})>>0$ of  $A_{\lambda}$ corresponding to  $\Psi(A_{\lambda})$.
    \item [ii] Assume that for each $\lambda \in [0, \Lambda^*]$, $\theta>0$
   $$
   f^{\pm}(\theta \nu^i_{\lambda}) \leq f'(0)\theta \nu^i_{\lambda}.
   $$
   \end{itemize}
\end{enumerate}
Let $$
\Phi(\lambda)=\frac{1}{\lambda} \Psi(A_{\lambda})> 0.
$$
According to Lemma \ref{lmeigen}, we can expect the graph of $\Phi$ as in Fig. \ref{fig4}. For the example in Section \ref{example},
$\Phi$ is a strictly convex function of $\lambda$ and, clearly satisfies Lemma \ref{lmeigen}.

Now we state Lemma \ref{lmeigen}, which is a analogous result in Weinberger \cite{Weinberger1978} and Lui \cite{Lui1989}.
However, due to the fact that $f'(0)$ is only quasi-positive and the elements of $A_{\lambda}$ are not necessarily log convex,
some of its proof here are different from Lui \cite{Lui1989}.  A similar result is included in \cite{HwangSystemPDE}.
A theorem on the convexity of the dominant eigenvalue of matrices
due to Cohen \cite{Cohen1981} is used to show that $\Psi(A_{\lambda})$ is convex function of $\lambda$. Lemma \ref{lmeigen} improves \cite[Theorem 4.2]{Weinberger2002-1} by eliminating the case (b) in \cite[Theorem 4.2]{Weinberger2002-1}.

\slm\mlabel{lmeigen}  Assume that $(H1)-(H2)$ hold. Then
\begin{enumerate}
  \item [(1)] $\Phi(\lambda) \to \infty$ as $\lambda \to 0;$
  \item [(2)] If $\Lambda^* =\infty$, $\Phi(\lambda) \to \infty$ as $\lambda \to \infty;$
  \item [(3)] $\Phi(\lambda)$ is decreasing as $\lambda=0^+;$
  \item [(4)]  $\Psi(A_{\lambda})$ is a convex function of $\lambda>0;$
  \item [(5)] $\Phi'(\lambda) $ changes sign at most once on $(0, \infty)$
  \item [(6)] $\Phi(\lambda)$ has the minimum $$
c^*=\inf_{ \lambda >0}\Phi(\lambda)>0
$$
at a finite $\Lambda_{c^*}>0$.
\item [(7)] For each $c > c^*$, there exist a positive $\Lambda_{c}<\Lambda_{c^*}$ and $\gamma \in (1,\min\{2, \frac{\Lambda_{c^*}}{\Lambda_c}\})$ such that
$$\Phi(\Lambda_{c})=c,\;\; \Phi(\gamma \Lambda_{c})<c.$$
That is $$\frac{1}{\Lambda_c}A_{\Lambda_c} \nu_{\Lambda_c}= \Phi(\Lambda_{c})\nu_{\Lambda_c}=c\nu_{\Lambda_c}$$
and
$$\frac{1}{\gamma\Lambda_c}A_{\gamma\Lambda_c} \nu_{\gamma\Lambda_c}= \Phi(\gamma\Lambda_{c})\nu_{\gamma\Lambda_c}<c\nu_{\gamma\Lambda_c}$$
where $\nu_{\Lambda_c}>>0,\nu_{\gamma\Lambda_c}>>0$ are positive eigenvectors
of $\frac{1}{\Lambda_c}A_{\Lambda_c}, \frac{1}{\gamma\Lambda_c}A_{\gamma\Lambda_c}$
corresponding to eigenvalues $\Phi(\Lambda_{c})$ and $\Phi(\gamma\Lambda_{c})$ respectively.
\end{enumerate}
\elm
\pf
 The proof of the convexity of $\Psi(A_{\lambda})$ is
similar to that in  Crooks \cite{Crooks1996} for matrices with positive off-diagonal elements. It is easily seen that
$\Psi(A_{\lambda})=\rho(A_{\lambda}+\alpha I)-\alpha$ is non-decreasing function of $\lambda>0$ (\cite[Theorem 8.1.18]{Horn}).
Further, a theorem on the convexity of the dominant eigenvalue of matrices due to Cohen \cite{Cohen1981} states that for
any positive diagonal matrices $D_1, D_2$ and $t \in (0,1)$,
$$
\Psi(tD_1+(1+t)D_2+f'(0)) \leq t\Psi(D_1+f'(0))+(1-t)\Psi(D_2+f'(0))
$$
as before, here $\Psi(A)$ is the principle eigenvalue of $A$. Now if $\alpha_1, \alpha_2 \in \mathbb{R}$ and $t \in (0,1)$,
$$(t\alpha_1+(1-t)\alpha_2)^2 \leq t\alpha_1^2+(1-t)\alpha_2^2.$$
This implies that
\begin{equation*}
\begin{split}
\Psi(A_{t\lambda_1+(1-t)\lambda_2})& = \Psi((t\lambda_1+(1-t)\lambda_2)^2D+f'(0))\\
                       &\leq \Psi(t\lambda_1^2D+(1-t)\lambda_2^2D+f'(0))\\
                       &\leq t\Psi(\lambda_1^2D+f'(0))\\
                       & \quad +(1-t)\Psi(\lambda_2^2D+f'(0))\\
                       &=t\Psi(A_{\lambda_1})+(1-t)\Psi(A_{\lambda_2})\\
\end{split}
\end{equation*}
Since $\Psi(A_{\lambda})$ is a simple root of the characteristic equation of an irreducible block,
it can be shown that $\Psi(A_{\lambda})$ is twice continuously differentiable on $\mathbb{R}$. Thus
$$\Psi''(\lambda)\geq 0$$
and a calculation shows
$$
[\lambda\Phi(\lambda)]'=\Psi'(\lambda)
$$
$$
\Phi'(\lambda)=\frac{1}{\lambda}[\Psi'(\lambda)-\Phi(\lambda)]
$$
and
$$
(\lambda^2\Phi'(\lambda))'=\lambda\Psi''(\lambda)\geq 0.
$$
As for (2), we  need to prove that
$\lim_{\lambda \to \infty}\frac{\Psi(A_{\lambda})}{\lambda}=\infty$. In fact, there exists an $\epsilon>0$ such that all diagonal elements of
$D-\epsilon I$ are strictly positive, then $\Psi\big(D-\epsilon I\big)>0$ and choose $\lambda$ large enough so that
\begin{equation*}
\begin{split}
\Psi(A_{\lambda})& =\Psi(D\lambda^2+f'(0)) \\
                       &=\Psi\big((D-\epsilon I)\lambda^2+ (\epsilon \lambda^2 I+f'(0))\big)\\
                       &\geq \Psi\big((D-\epsilon I)\lambda^2\big)\\
                       & = \lambda^2 \Psi\big(D-\epsilon I\big)
\end{split}
\end{equation*}
Thus $\lim_{\lambda \to \infty}\frac{\Psi(A_{\lambda})}{\lambda}=\infty$.  (6) is a consequence of (1)-(5). (7) is a direct consequence
of (1)-(6). It is just  the fact that $\nu_{\lambda}>>0$ is a eigenvector of $\frac{1}{\lambda}A_{\lambda}$ corresponding to eigenvalue $\Phi(A_{\lambda})$
for $\lambda=\Lambda_c$ and $\gamma \Lambda_c$.
\epf

We now recall results on the spreading speeds in  Weinberger, Lewis and Li \cite{Weinberger2002-1} and Lui \cite{Lui1989}.
While Theorem 4.1 \cite{Weinberger2002-1} holds for non cooperative reaction-diffusion systems, it does require that the reaction-diffusion system has a single speed.
In general, such a condition is very difficult to verify. In the same section, for cooperative systems, Theorem 4.2 in \cite{Weinberger2002-1}
provides sufficient conditions to have a single speed. The following theorem combines the results of Theorems 4.1 and 4.2 in \cite{Weinberger2002-1},
which can be a consequence of Theorems 3.1 and 3.2 for discrete-time recursions in Lui \cite{Lui1989}.

\sthm (Weinberger, Lewis and Li \cite{Weinberger2002-1})\label{th20} Assume $(H1)-(H2)$ hold and (\ref{eq1}) is \textbf{cooperative}.  Then the following statements are valid:
\begin{enumerate}
\item [(i)] For any $u_0 \in \mathcal{C}_{k}$ with compact support, let $u(x,t)$ be the solution of (\ref{eq1}) with (\ref{eq1bc}).
Then
$$
\lim_{t \to \infty}  \sup_{\abs{x}\geq ct} u(x,t) = 0, \text{ for } c> c^*
$$

\item [(ii)] For any strictly positive vector $\omega \in \mathbb{R}^N$, there is a positive $R_{\omega}$ with the property that if $u_0 \in \mathcal{C}_{k}$
and $u_0 \geq \omega $
on an interval of length $2R_{\omega}$, then the solution $u(x,t)$ of (\ref{eq1})  with (\ref{eq1bc}) satisfies
$$
\liminf_{t \to \infty} \inf_{\abs{x}\leq tc} u(x,t) =  k, \text{ for } 0< c< c^*
$$
\end{enumerate}
\ethm
In another paper \cite{Weinberger2005}, for cooperative systems, Li, Weinberger and Lewis established that the slowest spreading
speed $c^*$ can always be characterized as the slowest speed of a family of traveling waves.
These results describe the properties of spreading speed $c^*$ for monotone systems. Based on these spreading results for cooperative systems,
we will discuss analogous spreading speed results for non cooperative systems.

\section{Results on general non-cooperative systems}\label{rest}
In this section, we state a theorem for general partially cooperative reaction-diffusion systems (\ref{eq1})
which establishes the existence of traveling waves and  spreading speed for a large class of non-cooperative
systems. As we discussed in Section \ref{preli}, assumptions (H1-H2) and the proofs in Section \ref{spreadings} are different from those
in \cite{HwangSystemPDE} and the assumptions seem easier to verify.
Although the existence of traveling wave solutions for cooperative systems are known (see,e.g. \cite{Weinberger2005}), we shall prove the existence of
traveling wave solutions for both cooperative and non cooperative systems as our proofs for non cooperative systems are based on those for cooperative systems.
Further, in additions to the existence of traveling wave solutions, we shall be able to obtain asymptotic behavior of
the traveling wave solutions in terms of eigenvalues and eigenvectors for both cooperative and non cooperative systems.
The following theorem is our results for general non-cooperative reaction-diffusion systems.

\sthm\label{th30} Assume $(H1)-(H2)$ hold.  Then the following statements are valid:
\begin{enumerate}
\item [(i.)] For any $u_0 \in \mathcal{C}_{k}$ with compact support, the solution $u(x,t)$ of (\ref{eq1})  with (\ref{eq1bc}) satisfies
$$
\lim_{t \to \infty}  \sup_{\abs{x}\geq tc} u(x,t) = 0, \text{ for } c> c^*
$$

\item [(ii.)] For any vector $\omega \in \mathbb{R}^N, \omega>>0$, there is a positive $R_{\omega}$ with the property that if $u_0 \in \mathcal{C}_{k}$
and $u_0 \geq \omega $
on an interval of length $2R_{\omega}$, then the solution $u(x,t)$ of (\ref{eq1})  with (\ref{eq1bc}) satisfies
$$
k^- \leq \liminf_{t \to \infty} \inf_{\abs{x}\leq tc} u(x,t) \leq  k^+, \text{ for } 0< c< c^*
$$

\item [(iii.)] For each $c > c^*$ (\ref{eq1}) admits a traveling wave solution $u=u(x+ct)$ such that
$0 << u(\xi) \leq k^+, \xi \in \mathbb{R}$,
$$k^-\leq \liminf_{\xi \to \infty}u(\xi) \leq \limsup_{\xi \to \infty}u(\xi)\leq k^+$$
and
\begin{equation}\label{aystomth112}
\lim_{\xi \to -\infty} u(\xi)e^{-\Lambda_{c} \xi}=\nu_{\Lambda_{c}}.
\end{equation}
If, in addition, (\ref{eq1}) is cooperative in $\mathcal{C}_{k}$, then $u$ is nondecreasing on $\mathbb{R}$.

\item [(iv.)] For $c = c^*$ (\ref{eq1}) admits a nonconstant traveling wave solution $u=u(x+ct)$ such that
$0 \leq  u(\xi) \leq k, \xi \in \mathbb{R}$,
$$k^-\leq \liminf_{\xi \to \infty}u(\xi) \leq \limsup_{\xi \to \infty}u(\xi)\leq k^+.$$

\item [(v.)] For $0<c<c^*$ (\ref{eq1}) does not admit a traveling wave solution $u=u(x+ct)$
 with $ \liminf_{\xi \to \infty} u(\xi)>>0$ and $u(-\infty)=0.$
\end{enumerate}
\ethm

\srmark\label{rem1}
In many cases, $f^{\pm}$ can be taken as piecewise functions consisting of $f$ and appropriate constants as
demonstrated in Section \ref{example}. In order to have a better estimate
for the traveling wave solution $u$ for non cooperative systems, it is desirable to choose two function $f^{\pm}$ which are close enough.
The smallest monotone function above $f$ and the largest monotone function below $f$  are natural choices of $f^{\pm}$
if they satisfy other requirements, See \cite{Thieme1979,hsu2008,LiLewis2009} for the discussion for scalar cases and \cite{WeinbergerKS2009}
for a partially cooperative reaction-diffusion system. Our construct of $f^-$ in Section \ref{example} is different from the previous papers.
\ermark

\srmark\label{rem2}
When (\ref{eq1}) is cooperative in $\mathcal{C}_{k}$, then $f^{\pm}=f.$
\ermark

\srmark\label{rem3} Assumptions (H1)(i-ii) imply that  $\mathcal{C}_{k^{+}}$ is an invariant set of (\ref{eq1}) in the sense that
             for any given $u_0 \in \mathcal{C}_{k^+}$, the solution of (\ref{eq1}) with the initial condition $u_0$ exists and remains
             in $\mathcal{C}_{k^{+}}$ for $t \in [0,\infty)$.  In fact,
for a given $u_0 \in \mathcal{C}_{k^+}$, let $u(x,t)$ be the solution of (\ref{eq1})
with the initial condition $u_0$. Theorem \ref{comparison1} implies that
$$
0 \leq u(x,t) \leq k^+, x \in \mathbb{R}, t >0.
$$
Now according to Smoller \cite[Theorem 14.4]{Smoller1994} (\ref{eq1}) (and also (\ref{eq1+}), (\ref{eq1-})) has a solution $u$ for $t \in [0, \infty)$ and $0 \leq u \leq k^+$ if
the initial value $u_0$ is uniformly continuous on $\mathbb{R}$.
\ermark

We shall prove Theorem \ref{th30} (i)-(iii) in Section \ref{spreadings} and  (iii)-(v) in Section \ref{travel}.

\section{Results on a model arising from epidermal wound healing}\label{example}

In Section \ref{example}, we shall apply the general results in Section \ref{rest} to the model (\ref{eqoper00}) arising from epidermal wound healing.
This model is not cooperative because of the fact that $h(u_1)$ is not monotone. We shall establish the existence of traveling waves as well as the results on the speed of propagation to (\ref{eqoper00}). In addition,
we characterize  the spreading speed as the slowest speed of a family of non-constant traveling wave solutions of (\ref{eqoper00}).  The spreading speed for (\ref{eqoper00}) was discussed
in \cite{Sherratt1991,murray2003} based on  numerical methods and singular perturbation techniques for several special cases, for example, $d_1=0$.

Recall that $h(u_1)=\frac{u_1(1+\zeta^2)}{u_1^2+\zeta^2}$. It is easy to (\ref{eqoper00}) has two equilibria $(0,0)$ and $(1,1)$.
In fact, the following equalities hold at its non-trivial equilibrium
\begin{equation}\label{equilib2}
\begin{split}
h(u_1) &=\frac{\frac{1}{2-u_1}+\kappa-1}{\kappa} \\
u_2&=h(u_1).
\end{split}
\end{equation}
Now it is clear that (\ref{equilib2}) has only one positive solution $(1,1).$ In fact, $\frac{\frac{1}{2-u_1}+\kappa-1}{\kappa} $
in (\ref{equilib2}) is increasing and convex on $(0, \infty)$ and  the first equation of (\ref{equilib2})
has only one solution $u_1=1$.

We  now need to check (H2). The linearization of (\ref{eqoper00}) at the origin  is
\begin{equation}\label{eqoper10}
\begin{split}
\frac{\partial u_1}{ \partial t}&=d_1 \Delta u_1+(1-2\kappa)u_1\\
\frac{\partial u_2}{ \partial t}&=d_2 \Delta u_2+b(h'(0)u_1-u_2)
\end{split}
\end{equation}
where $h'(0)=\frac{1+\zeta^2}{\zeta^2}.$ The matrix $A_{\lambda}$ in (\ref{egenvalue}) for (\ref{eqoper00}) is
\begin{equation}\label{matrix17}
\begin{split}
A_{\lambda}=(a^{i,j}_{\lambda})=\left(
      \begin{array}{ll}
        d_1\lambda^2+1-2\kappa & \;\;\;0\\
        bh'(0)  & d_2 \lambda^2-b\\
      \end{array}
    \right)
 \end{split}
\end{equation}

It is easy to see that
\begin{equation}\label{eqg779}
\begin{split}
h(u_1) < h'(0)u_1, \;\; u_1 \in (0, \infty).
\end{split}
\end{equation}

In order to use Theorem \ref{th30}, we shall define the two monotone systems. Note that $h(u_1)$ achieves its maximum value $\frac{1+\zeta^2}{2\zeta}$
when $u_1=\zeta.$
\begin{equation*}
h^{+}(u_1) = \left\{ \begin{array}{ll}
h(u_1),    & \;\;\;\;\; 0 \leq u_1 \leq \zeta, \\[.2cm]
h(\zeta), & \;\;\;\;\;  u_1 \geq \zeta.
\end{array} \right.
\end{equation*}
and the corresponding cooperative system is
\begin{equation}\label{eqoper1}
\begin{split}
\frac{\partial u_1}{ \partial t}&=d_1 \Delta u_1+s(u_2)u_1(2-u_1)-u_1\\
\frac{\partial u_2}{ \partial t}&=d_2\Delta u_2+b(h^+(u_1)-u_2)\\
\end{split}
\end{equation}
In a similar manner, one can find (\ref{eqoper1}) has two equilibrium $(0,0)$ and  $(k_1^+, k_2^+)$ satisfying
\begin{equation}\label{equilib2+}
\begin{split}
h^+(k_1^+) &=\frac{\frac{1}{2-k_1^+}+\kappa-1}{\kappa} \\
k_2^+&=h^+(k_1^+).
\end{split}
\end{equation}
Since  $h^+ =  h$ for $u_1 \leq \zeta$, then $k_1^+ > \zeta$ (if $k_1^+ \leq  \zeta$ , then $k^+_1=1$ and $\zeta \geq 1$) and $h^+(k_1^+)= \frac{1+\zeta^2}{2\zeta}$. Solving $k_1^+$ directly from (\ref{equilib2+})
gives that $k_1^+= 2-\frac{1}{1+(\frac{1+\zeta^2}{2\zeta}-1)\kappa}>1>\zeta$. It follows that $k^+_2 = h^+(k_1^+)=\frac{1+\zeta^2}{2\zeta} >1$.

Now there is a $h_0 \in (0, \zeta]$ such that $h(h_0)<\min\{1,h(k_1^+)\}$ and define
\begin{equation*}
h^{-}(u_1) = \left\{ \begin{array}{ll}
h(u_1),    &\;\;\; 0 \leq u_1 \leq h_0, \\[.2cm]
h(h_0), & \;\;\; u_1 > h_0.
\end{array} \right.
\end{equation*}
It is clear that
$$0<h^-(u_1) \leq h(u_1) \leq  h^+(u_1) \leq h'(0)u_1, u_1 \in (0, k^+_1]$$
and $h^-(u_1) <1$ for $u_1\geq 0.$

\begin{figure}
\begin{center}
  \includegraphics[width=8cm]{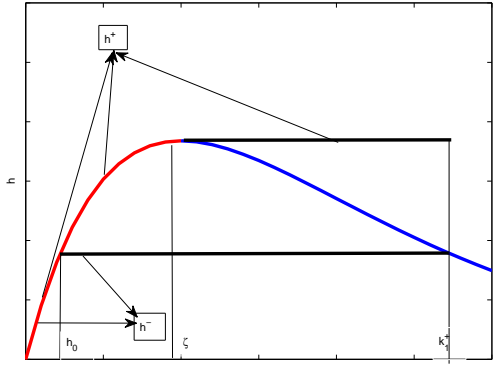}
   \caption{The construction of $h^+$ and $h^-$. The red curve is $h$.}\label{fig1}
\end{center}
\end{figure}

The corresponding cooperative system for $h^-$ is
\begin{equation}\label{eqoper2}
\begin{split}
\frac{\partial u_1}{ \partial t}&=d_1 \Delta u_1+s(u_2)u_1(2-u_1)-u_1\\
\frac{\partial u_2}{ \partial t}&=d_2\Delta u_2+b(h^-(u_1)-u_2)\\
\end{split}
\end{equation}
In a similar manner, one can find (\ref{eqoper2}) has two equilibrium $(0,0)$ and  $(k_1^-, k_2^-)$ satisfying
\begin{equation}\label{equilib2-}
\begin{split}
h^-(k_1^-) &=\frac{\frac{1}{2-k_1^-}+\kappa-1}{\kappa} \\
k_2^-&=h^-(k_1^-).
\end{split}
\end{equation}
Since  $h^- =  h$ for $u_1 \leq h_0$, then $k_1^- > h_0$ and $h^-(k_1^-)= h(h_0)$. Solving $k_1^-$ directly from (\ref{equilib2-})
gives that $k_1^- =2-\frac{1}{1+(h(h_0)-1)\kappa}<1$ as $h(h_0)<1$. On the other hand, because $h(h_0)>0$, a simple calculation shows that
$k_1^-> \frac{1-2\kappa}{1-\kappa}>0$. As before we have $0<k_2^- = h^-(k_1^-)=h(h_0) <1$.

Thus, $$
(0,0) <<(k_1^-, k_2^-) \leq (1, 1) \leq (k_1^+, k_2^+).
$$
We can always extend (\ref{eqoper1}),(\ref{eqoper2}) to be  Liptschz continuous in $\mathbb{R}^2$ without changing the functions in the region
$[0, k^+]$.  Then Theorem \ref{comparison1} implies that
$$
0 \leq u(x,t) \leq k^+, x \in \mathbb{R}, t >0.
$$
Thus we are only interested in the invariant region. Now it is straightforward to check all other conditions of
(H1)(i)-(iii).

The spreading results for the cooperative systems were used to establish
in \cite{WeinbergerKS2009}. We now demonstrate Theorem \ref{th30} can be used to establish spreading speed and
traveling wave solutions of the nonmonotone system (\ref{eqoper00}) and summarize our main results in the following Theorem.

\sthm\label{th33}  Let $ d_1, d_2$ be all positive numbers and $\kappa \in (0, \frac{1}{2}), \zeta \in (0,1)$,
\begin{equation}\label{eqg-141}
\frac{d_2}{d_1} < 2 + \frac{b}{1-2\kappa}.
\end{equation}
and
\begin{equation}\label{eqg-142}
\frac{2\kappa h'(0)}{1-\kappa} \leq  \left\{\begin{array}{cc}
\displaystyle 1+ \frac{1-2\kappa}{b} & d_1\geq d_2 , \\
\displaystyle (2-\frac{d_2}{d_1})\frac{1-2\kappa}{b}+1, &  d_1\leq d_2 \\
\end{array}\right.
\end{equation}
Then  the conclusions of Theorem \ref{th30} hold for (\ref{eqoper00}) where the minimum speed $c^*=2\sqrt{(1-2\kappa)d_1}$,
$\Lambda_c=\frac{c-\sqrt{c^2-4d_1(1-2\kappa)}}{2d_1}>0$ and  $\nu_{\Lambda_c}$ is defined in (\ref{egenvectro4907}).
That is, the solution $(u_1(x,t), u_2(x,t)$ of (\ref{eqoper00}) satisfies
\begin{enumerate}
\item [(i.)] If the functions $(u_1(x,0), u_2(x,0))\leq (k_1,k_2)$ are nonnegative continuous and have compact support, then
$$
\lim_{t \to \infty}  \sup_{\abs{x}\geq tc} (u_1(x,t), u_2(x,t)) = (0,0) \text{ for } c> c^*
$$

\item [(ii.)] If the functions $(u_1(x,0), u_2(x,0))\leq (k_1,k_2)$ are  nonnegative continuous  and $u_1(x,0) \not\equiv 0$, then
$$
(k^-_1, k^-_2) \leq \liminf_{t \to \infty} \inf_{\abs{x}\leq tc} (u_1(x,t), u_2(x,t)) \leq  (k^+_1, k^+_2), \text{ for } 0< c< c^*
$$

\item [(iii.)] For each $c > c^*$ (\ref{eqoper00}) admits a traveling wave solution $(u_1(\xi), u_2(\xi))$ such that
$(0,0) << (u_1(\xi), u_2(\xi)) \leq (k^+_1, k^+_2), \xi \in \mathbb{R}$,
$$(k^-_1, k^-_2)\leq \liminf_{\xi \to \infty} (u_1(\xi), u_2(\xi)) \leq \limsup_{\xi \to \infty}(u_1(\xi), u_2(\xi))\leq (k^+_1, k^+_2)$$
and
\begin{equation}\label{aystomth112-1}
\lim_{\xi \to -\infty} (u_1(\xi), u_2(\xi)) e^{-\Lambda_{c} \xi}=\nu_{\Lambda_{c}}.
\end{equation}
\item [(iv.)] For $c = c^*$ (\ref{eq1}) admits a nonconstant traveling wave solution $(u_1(\xi), u_2(\xi))$ such that
$(0,0) << (u_1(\xi), u_2(\xi)) \leq (k^+_1, k^+_2), \xi \in \mathbb{R}$,
$$(k^-_1, k^-_2)\leq \liminf_{\xi \to \infty} (u_1(\xi), u_2(\xi))\leq \limsup_{\xi \to \infty}u(\xi)\leq (k^+_1, k^+_2).$$

\item [(v.)] For $0<c<c^*$ (\ref{eq1}) does not admit a traveling wave solution $(u_1(\xi), u_2(\xi))$
 with $ \liminf_{\xi \to \infty} (u_1(\xi), u_2(\xi)) >>(0,0)$ and $(u_1(-\infty), u_2(-\infty))=0.$
\end{enumerate}
\ethm

Now we shall verify (H2) for (\ref{eqoper1}). In fact, the principle eigenvalue of $A_{\lambda}$ is $\Psi(A_{\lambda})=d_1\lambda^2+1-2\kappa$,
which is a convex function of $\lambda$. Furthermore, $$\Phi(\lambda)=\frac{\Psi(A_{\lambda})}{\lambda}=\frac{d_1\lambda^2+1-2\kappa}{\lambda}$$
satisfies the results of Lemma  \ref{lmeigen}. In fact $\Phi(\lambda)$ is  also a strictly convex function of $\lambda$.
The minimum of $\Phi(\lambda)$ is $c^*=2\sqrt{(1-2\kappa)d_1}$ when $\lambda =\frac{\sqrt{1-2\kappa}}{\sqrt{d_1}}$. For each $c \geq c^*$, the left positive solution of $\Phi(\lambda)=c$
 in Lemma \ref{lmeigen} is
 $$\Lambda_c=\frac{c-\sqrt{c^2-4d_1(1-2\kappa)}}{2d_1}$$.

In particular, $$\Lambda_{c^*}=\frac{\sqrt{1-2\kappa}}{\sqrt{d_1}}.$$
For each $0 \leq \lambda \leq \Lambda_{c^*}$, the positive eigenvector of $A_{\lambda}$
corresponding to $\Psi(A_{\lambda})$ is
\begin{equation}\label{egenvectro4907}
\nu_{\lambda}= \left(
                     \begin{array}{c}
                       \nu_{\lambda}^{1} \\
                       \nu_{\lambda}^{2} \\
                     \end{array}
                   \right)
=\left(\begin{array}{c}
                                                                 (d_1-d_2)\lambda^2+1-2\kappa+b\\
                                                                                         bh'(0)
                                                                                                        \end{array}
                                                                                                        \right)
\end{equation}
 Because of (\ref{eqg-141}), $\nu_{\lambda}$ is a strictly positive vector for $ \lambda \in [0, \Lambda_{c^*}].$
This is clear when $d_1 \geq d_2$. If $d_1 < d_2$ and (\ref{eqg-141}) holds, for $0 \leq \lambda \leq \Lambda_{c^*}$, we have
$$(d_1-d_2)\lambda^2+1-2\kappa+b \geq (d_1-d_2)\Lambda_{c^*}^2+1-2\kappa+b>0.$$

Further from (\ref{egenvectro4907}) we can see that
$$
\frac{\nu_{\lambda}^{2}}{\nu_{\lambda}^{1}}=\frac{bh'(0)}{(d_1-d_2)\lambda^2+1-2\kappa+b}=\frac{h'(0)}{\sigma}
$$
where $\sigma= 1+ \frac{(d_1-d_2)\lambda^2+1-2\kappa}{b}.$ For $\lambda\in [0, \Lambda_{c^*}]$, it is clear that
$\sigma> 1+ \frac{1-2\kappa}{b}$ if $d_1\geq d_2$ and $\sigma \geq (2-\frac{d_2}{d_1})\frac{1-2\kappa}{b}+1>0$
if $d_1<d_2$ and (\ref{eqg-141}) holds.

Let
$$
(u_1,u_2)=(\theta,\theta \frac{h'(0)}{\sigma})>>(0,0),\; \theta >0.
$$
Thus (H2)(ii) for (\ref{eqoper1}) is equivalent to the following two inequalities
\begin{equation}\label{eqoper12290}
\begin{split}
(\kappa u_2 +1 - \kappa)u_1(2-u_1)-u_1& \leq (1-2\kappa)u_1\\
b h(u_1)-bu_2& \leq bh'(0)u_1-bu_2.
\end{split}
\end{equation}
Because $h(u_1) \leq h'(0)u_1,$ (\ref{eqoper12290}) is equivalent to the following inequality
\begin{equation}\label{eqoper122}
\begin{split}
(2-u_1) \kappa & \leq (1-\kappa)\frac{u_1}{u_2}
\end{split}
\end{equation}
and the following inequality suffices to verify (\ref{eqoper122}) $$
\frac{2\kappa}{1-\kappa} \leq \frac{\sigma}{h'(0)}.
$$
which is true with (\ref{eqg-142}) because of the estimates for $\sigma$ for $d_1 \geq d_2$ and $d_1 < d_2$.
Notice that $h^{\pm}$ and $h$ are identical around the origin.  By the exact same arguments (just replacing $k_i^+$ by $k^-_i$ and $h^{+}$ by $h^{-}$),
 we can verify that (H2) holds for (\ref{eqoper2}) as well.

It remains to show that the condition (ii) in Theorem \ref{th30} can be satisfied if $u_1(x,0) \not\equiv 0.$  The arguments here is the same as
in Weinberger, Kawasaki and Shigesada \cite{WeinbergerKS2009}.  We choose positive constants $\rho, \eta$ so small that
\begin{equation}\label{constants}
-d_1\rho^2+(1-2\kappa)-(1-\kappa)\eta>0.
\end{equation}
Since $u_2(x,t) \geq 0$, we have
\begin{equation}\label{equforu1}
\begin{split}
\frac{\partial u_1}{ \partial t} & \geq d_1 \Delta u_1+(1-\kappa)u_1(2-u_1)-u_1\\
&=d_1 \Delta u_1+u_1\big((1-2\kappa)-(1-\kappa)u_1\big)
\end{split}
\end{equation}
By the strong maximum principle we have $u_1(x,t)>0$ for $t>0$.  Thus we can require that $\eta \leq u_1(x,t_1)$
for some $t_1>0$ and $|x| \leq \frac{\pi}{2\rho}$ and some $t_1>0$ by choosing $\eta$ small enough.  If $(\hat{u_1}, \hat{u_2})$ is
the solution of
\begin{equation}\label{eqoper21}
\begin{split}
\frac{\partial u_1}{ \partial t}&=d_1 \Delta u_1+u_1\big((1-2\kappa)-(1-\kappa)u_1\big)\\
\frac{\partial u_2}{ \partial t}&=d_2\Delta u_2+b(h^-(u_1)-u_2)\\
\end{split}
\end{equation}
with the initial values
\begin{equation}\label{initials}
u_1(x,t_1)=  \left\{\begin{array}{cc}
\displaystyle \eta \cos(\rho x) & \;\; \text{ for }\abs{x} \leq \frac{\pi}{2\rho} \\
\displaystyle 0 &  \;\; \text{ for } \abs{x} \geq \frac{\pi}{2\rho} \\
\end{array}\right.,\;\;\
u_2(x,t_1)=0.
\end{equation}
It is clear that (\ref{eqoper21}) has two equilibriums $(0,0)$ and $(\frac{1-2\kappa}{1-\kappa}, h^-(\frac{1-2\kappa}{1-\kappa}))>>0.$  Furthermore,
there is no other stationary solution of (\ref{eqoper21}) between the two equilibriums.

The comparison principle shows that the components of $(\hat{u_1}, \hat{u_2})$ are lower
bounds for $(u_1, u_2)$ when $t \geq t_1$. The inequality (\ref{constants}) shows that
both $( \frac{\partial \hat{u_1}}{\partial t}, \frac{\partial \hat{u_2}}{\partial t})$
are nonnegative at $t = t_1$, and the comparison principle then implies that $(\hat{u_1}, \hat{u_2})$ are nondecreasing in t.
It follows that $(\hat{u_1}, \hat{u_2})$ monotonically converges to $(\frac{1-2\kappa}{1-\kappa}, h^-(\frac{1-2\kappa}{1-\kappa}))$
uniformly in x on every bounded x-interval. Because $ (u_1, u_2)\geq (\hat{u_1}, \hat{u_2})$,
it follows that if we choose two positive constants $(\omega_1,\omega_1) $ with $(\omega_1,\omega_1)  < (\frac{1-2\kappa}{1-\kappa}, h^-(\frac{1-2\kappa}{1-\kappa}))$
then for all sufficiently large t, the condition (ii) in Theorem \ref{th30} is automatically satisfied on the fixed
interval $\abs{x} \leq 2R_{\omega} $. We thus obtain the statement (4.21) without an extra condition.

\begin{figure}
\begin{center}
  \includegraphics[width=5cm]{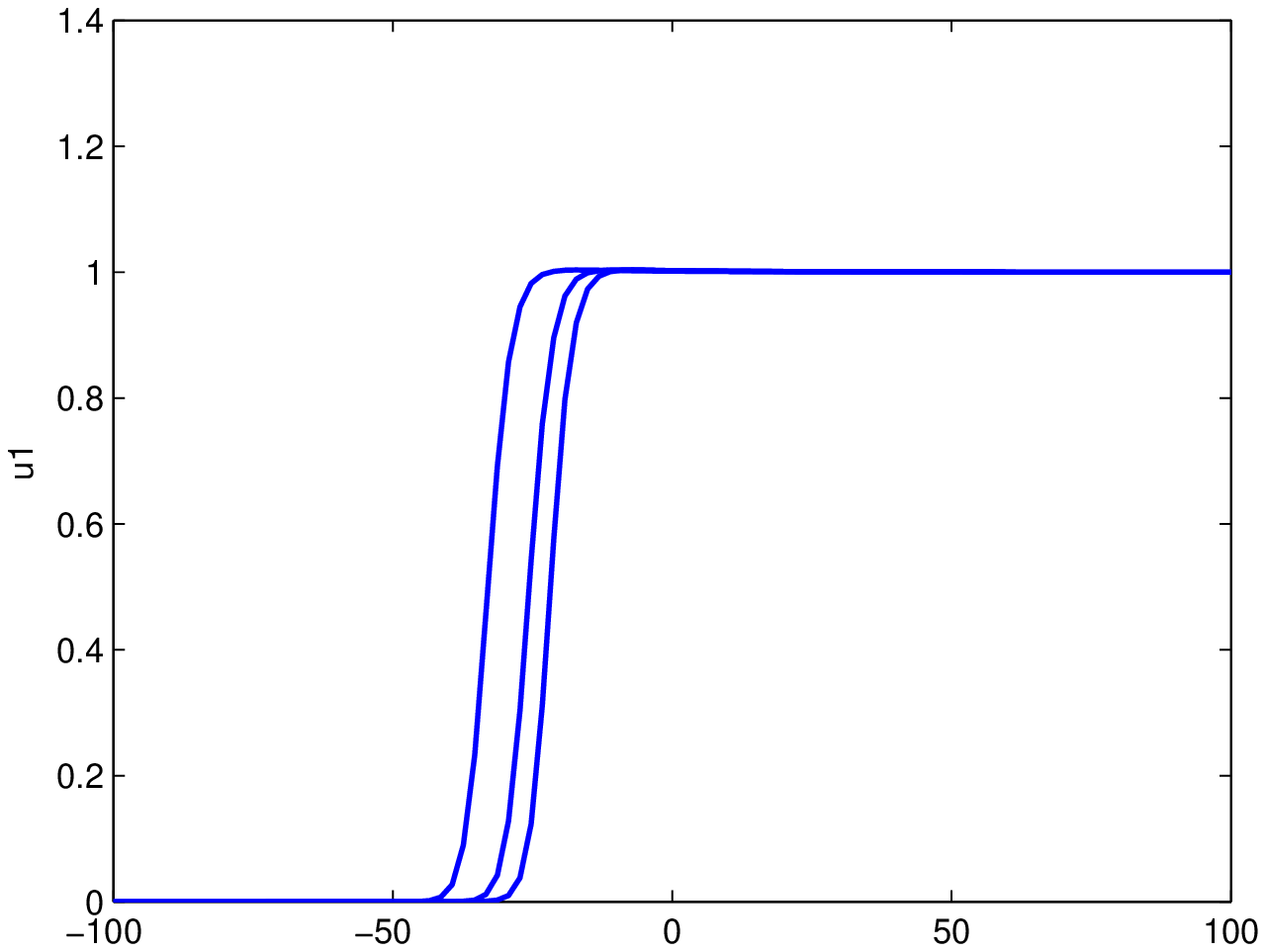} \includegraphics[width=5cm]{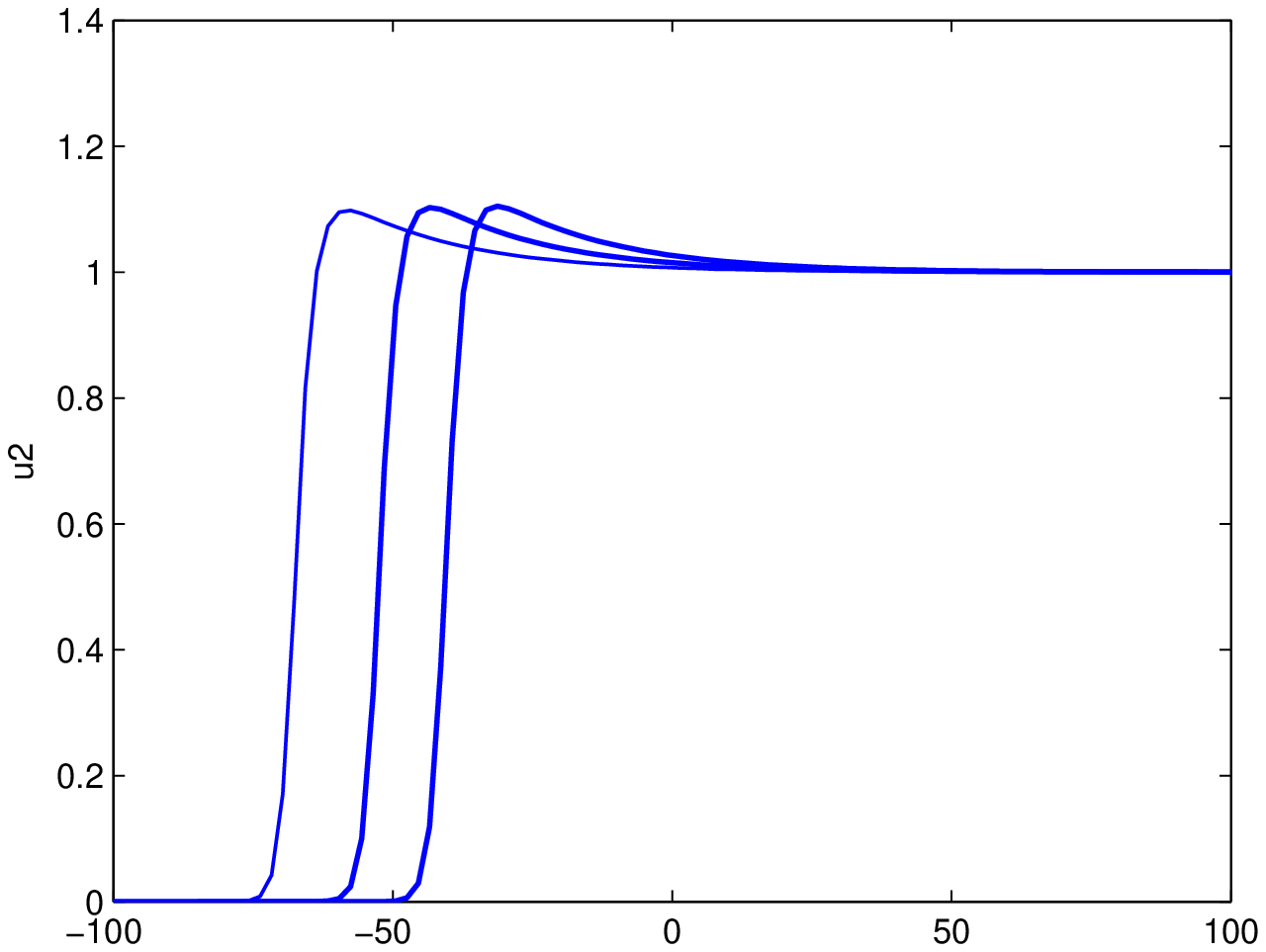}
  \caption{The simulations of the traveling wave solutions of (\ref{eqoper00}). We choose $d_1=d_2=1, \kappa=0.05, \zeta=0.1, b=0.05$.}\label{fig2}
\end{center}
\end{figure}

Fig. \ref{fig2} are the simulations of the traveling wave solutions of (\ref{eqoper00}). We choose $d_1=d_2=1, \kappa=0.05, \zeta=0.1, b=0.05$, which
satisfy the conditions of Theorem \ref{th33}. Note that the traveling solutions are not monotone and the minimum speed $c^*=2\sqrt{(1-2\kappa)d_1}=1.89.$

\section{The Spreading Speed}\label{spreadings}

\subsection{Comparison Principle}\label{comparsion}
We state the following comparison theorem for reaction-diffusion equations in  File \cite{Fife1979}.
The comparison principle is a consequence of the maximum principle (see, e.g., Protter and Weingberger \cite{Protter1984}).
Another form of the comparison principle from \cite{WeinbergerKS2009} was used in \cite{HwangSystemPDE}. As a result,
the proof in Section \ref{spreadings} is somewhat
different from \cite{HwangSystemPDE}.  By a suitable modification of the functions,
the conditions in this paper seem easier to verify.

%

\sthm \label{comparison1} Let $D$ be a positive definite diagonal matrix and $F=(F_j)$ vector-valued continuous functions in $\mathbb{R}^N$.
Assume that $F^{\pm}=(F_j^{\pm})$ are Lipschitz continuous on $\mathbb{R}$ and $F_j^{\pm}$ is nondecreasing in all but the $j$th component, $j=1,...,N$, and satisfying $$
F^{+}(u) \geq F(u) \geq F^{-}(u), u \in \mathbb{R}^N.
$$ Let $u, u^{\pm}$ be
continuous functions from $\mathbb{R} \times (0, T)$ into $\mathbb{R}^N$, $C^2$ in $\mathbb{R} \times (0, T)$, bounded and satisfying, for $i=1,...,n$
\begin{equation}\label{comp1}
\begin{split}
u_t-Du_{xx} &=  F(u)\\
u^{+}_t-Du^{+}_{xx} &\geq  F^{+}(u^{+})\\
u^{-}_t-Du^{-}_{xx} &\leq  F^{-}(u^{-})
\end{split}
\end{equation}
If $u^{+}(x, t_0) \geq u(x,t_0) \geq u^{-}(x,t_0), \;\; x\in \mathbb{R}$, then
$$
u^{+}(x,t) \geq u(x,t) \geq u^{-}(x,t), \;\; x\in \mathbb{R}, t \in [ t_0, T].
$$
\ethm


We are now able to prove Parts (i) and (ii) of Theorem \ref{th30}.

\subsection{Proof of Parts (i) and (ii) of Theorem \ref{th30}}

Part (i).  For a given $u_0 \in \mathcal{C}_{k}$ with compact support, let $u^+(x,t)$ be the solutions of (\ref{eq1+}) with the same initial
condition $u_0$, then Theorem \ref{comparison1} implies that $u^+(x,t) \in \mathcal{C}_{k^+}$ and
$$
0 \leq u(x,t) \leq u^{+}(x,t), x \in \mathbb{R}, t >0.
$$
Thus for any $c>c^*$, it follows from Theorem \ref{th20} (i) that $$
\lim_{t \to \infty} \sup_{\abs{x}\geq tc} u^{+}(x,t) = 0,
$$
and hence
$$
\lim_{t \to \infty} \sup_{\abs{x}\geq tc} u(x,t) = 0,
$$
Part (ii). According to  Theorem \ref{th20} (ii), for any strictly positive constant $\omega$, there is a positive $R_{\omega}$
(choose the larger one between the $R_{\omega}$ for (\ref{eq1+}) and the $R_{\omega}$ for (\ref{eq1-})) with the property that if $u_0 \geq \omega $
on an interval of length $2R_{\omega}$, then the solutions $u^{\pm}(x,t)$ of (\ref{eq1+}) and (\ref{eq1-}) with the same initial value $u_0$
are in $\mathcal{C}_{k^+}$ and satisfy
$$
\liminf_{t \to \infty} \inf_{\abs{x}\leq tc} u^{\pm}(x,t) =  k^{\pm}, \text{ for } 0< c< c^*.
$$
Thus, Theorem \ref{comparison1} implies that

$$
u^-(x,t) \leq u(x,t) \leq u^{+}(x,t), x \in \mathbb{R}, t >0.
$$
Thus for any $c<c^*$, it follow from Theorem \ref{th20} (ii) that $$
\liminf_{t \to \infty} \inf_{\abs{x}\leq ct} u^{\pm}(x,t) = k^{\pm},
$$
and hence
$$
k^- \leq \liminf_{t \to \infty} \inf_{\abs{x}\leq ct} u(x,t) \leq  k^+.
$$
\epf

\section{The characterization of $c^*$ as the slowest speeds of traveling waves}\label{travel}

\subsection{Equivalent integral equations and their upper and lower solutions}\label{upperlower}
In order to establish the existence of travel wave solutions, we fist set up equivalent integral equations.
Similar equivalent integral equations were also used before, see for example,  Wu and Zou \cite{wu2001}, Ma \cite{ma2001,ma2007} and the author \cite{Hwang2009}.
For the convenience of analysis, in this paper and \cite{Hwang2009}, both $\lambda_{1i}, \lambda_{2i}$ are chosen to be positive, and $-\lambda_{1i}, \lambda_{2i}$
are solutions of (\ref{characteis}).

Let $\beta>\max\{\abs{\partial_i f_j(x)}, x \in [0,k^+], i,j=1,...,N\} > 0.$
For $c>c^*$, the two solutions of the following equations,
\begin{equation}\label{characteis}
d_i\lambda^2 -c\lambda -\beta=0, i=1,...,N
\end{equation}
are
$-\lambda_{1i}$ and $\lambda_{2i}$ where
$$
\lambda_{1i}= \frac{-c + \sqrt{c^2+4\beta d_i}}{2d_i}>0, \lambda_{2i}= \frac{c+\sqrt{c^2+4\beta d_i}}{2d_i}>0.
$$
We choose $\beta$ sufficiently large so that
\begin{equation}\label{eq14-1}
\lambda_{2i}>\lambda_{1i} > 2\Lambda_{c}, i=1,...,N.
\end{equation}
Let $u=(u_i) \in \mathcal{C}_{k}$ and define a operator $\mathcal{T}[u]=(\mathcal{T}_i[u])$ by
\begin{equation}\label{eq3}
\begin{split}
\mathcal{T}_i[u](\xi) &= \frac{1}{d_i(\lambda_{1i}+\lambda_{2i})}\Big(\int_{\infty}^{\xi}e^{-\lambda_{1i}(\xi-s)}H_i(u(s))ds\\
&\quad +\int^{\infty}_{\xi}e^{\lambda_{2i}(\xi-s)}H_i(u(s))ds\Big)
\end{split}
\end{equation}
where
$$H_i(u(s))= \beta u_i(s)+ f_i(u(s)),$$
$\mathcal{T}_i[u], i=1,...,N$ is defined on $\mathbb{R}$ if $H_i(u), i=1,2$ is a bounded continuous function. In fact, the following identity
holds
\begin{equation}\label{eq13}
\begin{split}
&\frac{1}{d_i(\lambda_{1i}+\lambda_{2i})}\big (\int_{-\infty}^{\xi}e^{-\lambda_{1i}(\xi-s)} \beta ds +\int^{\infty}_\xi e^{\lambda_{2i}(\xi-s)}\beta ds \big )\\
&=\frac{\beta }{d_i(\lambda_{1i}+\lambda_{2i})} \big (\frac{1}{\lambda_{1i}} + \frac{1}{\lambda_{2i}}\big )= \frac{\beta }{d_i(\lambda_{1i} \lambda_{2i})}\\
&=1.
\end{split}
\end{equation}

In fact, a fixed point $u$ of $\mathcal{T}$ or solution of the equation
\begin{equation}\label{eq3fixpoint}
u(\xi)= \mathcal{T}[u](\xi)\;\; \xi \in \mathbb{R},
\end{equation}
is a traveling wave solution of (\ref{eq1}) in Lemma \ref{verify}. Lemma \ref{verify} summarizes the conclusion, which can be verified
in the same manner as in \cite{Hwang2009} for scalar cases by directly substituting derivatives of $u(\xi)$
into (\ref{eq211}). We  omit its proof here.

\slm\label{verify} Assume $(H1-H2)$ hold.
If $u\in \mathcal{C}_{k}$ is a fixed point of $\mathcal{T}[u]$, $$u(\xi)= \mathcal{T}[u](\xi)\;\; \xi \in \mathbb{R},$$
then
$u$ is a solution of (\ref{eq211}).
\elm
We now define upper and lower solutions of (\ref{eq3fixpoint}), $\phi^+$ and $\phi^-$, which are only continuous on $\mathbb{R}$.
Similar upper and lower solutions have been frequently used in the literatures. See Diekmann \cite{Diekmann1978JMB},
Weinberger \cite{Weinberger1978}, Liu \cite{Lui1989}, Weinberger, Lewis and Li \cite{Weinberger2002-1}, Rass and  Radcliffe \cite{Rass2003},
Weng and Zhao \cite{Weng2006} and more recently, Ma \cite{ma2007}, Fang and Zhao \cite{Fang2009} and Wang \cite{Hwang2009,HwangIntegralDiff}.
In particular, it is believed that the vector-valued lower solutions of the form in this paper  first appeared in \cite{Weng2006} for multi-type SIS epidemic models. In this paper, the upper and lower solutions here are defined and verified for
the integral operator other than differential equations.

\strdef\label{defupper}
A bounded continuous function $u=(u_i) \in C(\mathbb{R}, [0, \infty)^N)$  is an upper solution of (\ref{eq3fixpoint}) if
$$
\mathcal{T}_i[u](\xi) \leq u_i(\xi), \;\; \text{for all } \xi \in \mathbb{R}, i=1,...,N;
$$
a bounded continuous function $u=(u_i) \in C(\mathbb{R}, [0, \infty)^N)$ is a lower solution of (\ref{eq3fixpoint}) if
$$
\mathcal{T}_i[u](\xi) \geq u_i(\xi), \;\; \text{for all } \xi \in \mathbb{R}, i=1,...,N.
$$
\eeddef

Let $c>c*$ and consider the positive eigenvalue $\Lambda_{c}$ and corresponding eigenvector $\nu_{\Lambda_c}=(\nu^i_{\Lambda_c})$, $\gamma$
in Lemma \ref{lmeigen} and $q>1.$
Define
$$
\phi^+(\xi)=(\phi^+_i),
$$
where
$$
\phi^+_i=\min\{k_i, \nu^i_{\Lambda_c}e^{\Lambda_{c}\xi}\}, i=1,...,N,\; \xi \in \mathbb{R};
$$
and
$$
\phi^-(\xi)=(\phi^-_i),
$$
$$
\phi^-_i=\max\{0, \nu^i_{\Lambda_{c}}e^{\Lambda_{c}\xi}-q\nu^i_{\gamma \Lambda_{c}}e^{\gamma\Lambda_{c}\xi}\},i=1,...,N,\; \xi \in \mathbb{R}.
$$
%

It is clear that if $\xi \geq  \frac{\ln \frac{k_i}{\nu^i_{\Lambda_{c}}}}{\Lambda_{c}}$, $\phi^+_i(\xi)=k_i$, and
$\xi < \frac{\ln \frac{k_i}{\nu^i_{\Lambda_{c}}}}{\Lambda_{c}}$, $\phi^+_i(\xi)=\nu^i_{\Lambda_{c}}e^{\Lambda_{c}\xi},i=1,...,N.$

Similarly, if  $\xi \geq \frac{\ln( q \frac{\nu^i_{\gamma \Lambda_{c}}}{\nu^i_{\Lambda_{c}}}) }{(1-\gamma)\Lambda_{c}}$, $\phi^-_i(\xi)=0$,
and for $\xi < \frac{\ln( q \frac{\nu^i_{\gamma \Lambda_{c}}}{\nu^i_{\Lambda_{c}}}) }{(1-\gamma)\Lambda_{c}}$,
$$\phi^-_i(\xi)=\nu^i_{\Lambda_{c}}e^{\Lambda_{c}\xi}-q\nu^i_{\gamma \Lambda_{c}}e^{\gamma\Lambda_{c}\xi}, i=1,...,N.$$
We choose $q>1$ large enough that
$$
\frac{\ln( q \frac{\nu^i_{\gamma \Lambda_{c}}}{\nu^i_{\Lambda_{c}}}) }{(1-\gamma)\Lambda_{c}}<\frac{\ln \frac{k_i}{\nu^i_{\Lambda_{c}}}}{\Lambda_{c}}, i=1,...,N
$$
and then
$$
\phi^+_i(\xi) > \phi^-_i(\xi), i=1,...,N, \xi \in \mathbb{R}.
$$

\begin{figure}
\begin{center}
  \includegraphics[width=3cm]{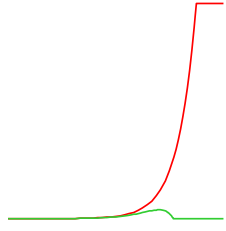}\\
  \caption{For each i, the curve above is $\phi^+_i$ and the below is $\phi^-_i$.}\label{upperlowerslu}
\end{center}
\end{figure}

We now prove that $\phi^+$ and $\phi^-$ are upper and lower solution of (\ref{eq3fixpoint}) respectively.

\slm\label{upper} Assume $(H1)-(H2)$ hold and (\ref{eq1}) is cooperative.  For any $c > c^*$, $\phi^+$ defined above is an upper solution of (\ref{eq3fixpoint}).
\elm
\pf
Note that when (\ref{eq1}) is cooperative, $f^{\pm}=f$ (see Remark \ref{rem2}). Let $\xi^*_i= \frac{\ln \frac{k_i}{\nu^i_{\Lambda_{c}}}}{\Lambda_{c}}, i=1,...,N.$ We first need to verify that $u=(u_i)=\phi^+$ is an upper solution of (\ref{eq211}) in the sense that
\begin{equation}\label{uppereq211}
d_iu''_i(\xi)-cu_i'(\xi)+f_i(u(\xi)) \leq 0, \;\;, i=1,...,N,\; \xi \neq  \xi^*_i.
\end{equation}
In view of the fact that $f_i(u(\xi)) \leq  f_i(k)=0$ for $\xi > \xi^*_i$, it is clear that (\ref{uppereq211}) holds for $\xi > \xi^*_i$.
For  $\xi < \xi^*_i$, $u_i(\xi)=\nu^i_{\Lambda_{c}}e^{\Lambda_{c}\xi}$.  From (H1-H2), we have, for  $\xi < \xi^*_i$ and $i=1,...,N$
\begin{equation}\label{uppereq40}
\begin{split}
& d_iu_i''(\xi)-cu_i(\xi)+f_i(u(\xi))\\
&\leq  d_i \Lambda_{c}^2\nu^i_{\Lambda_{c}}e^{\Lambda_{c}\xi}-c \Lambda_{c}\nu^i_{\Lambda_{c}}e^{\Lambda_{c}\xi}+\sum_{j=1}^n\partial_j f_i(0)\nu^j_{\Lambda_{c}}e^{\Lambda_{c}\xi}\\
& =\sum_{j=1}^na^{ij}_{\Lambda_{c}}\nu^j_{\Lambda_{c}}e^{\Lambda_{c}\xi}-c \Lambda_{c}\nu^i_{\Lambda_{c}}e^{\Lambda_{c}\xi}\\
&=\Lambda_{c} \Phi(\Lambda_{c})\nu^i_{\Lambda_{c}}e^{\Lambda_{c}\xi}-c \Lambda_{c}\nu^i_{\Lambda_{c}}e^{\Lambda_{c}\xi}\\
&=0.
\end{split}
\end{equation}
Since $u'_i(\xi^*-) \geq u'_i(\xi^*+)=0$, the proof of Lemma 2.5 in Ma \cite{ma2001} immediately implies that
$$
\mathcal{T}_i[u](\xi) \leq u_i(\xi), \;\; \text{for all } \xi \in \mathbb{R}, i=1,...,N.
$$
Thus $u=(u_i)=\phi^+$ is an upper solution of  (\ref{eq3fixpoint}).
Note that the upper solutions in this paper are defined for equivalent integral equations other than differential equations in \cite{ma2001}.
\epf
We now need the following estimate on $f$, which is an application of the Taylor's Theorem for multi-variable functions. Also see \cite{Hwang2009}.

\slm\label{estimageg} Assume $(H1-H2)$ hold. There exist positive constants $b_{ij}, i,j=1,...,N$ such that
$$
f_i(u)\geq \sum_{j=1}^N\partial_j f_i(0)u_j- \sum_{j=1}^Nb_{ij}(u_j)^2, \;\; u=(u_i), u_i \in [0,k_i], i=1,...,N.
$$
\elm

\slm\label{sub} Assume $(H1)-(H2)$ hold. For any $c > c^*$, $\phi^-$ defined above is a lower solution  of (\ref{eq3fixpoint})
if  $q$ (which is independent of $\xi$) is sufficiently large.
\elm
\pf
Let Let $\xi^*_i=\frac{\ln( q \frac{\nu^i_{\gamma \Lambda_{c}}}{\nu^i_{\Lambda_{c}}}) }{(1-\gamma)\Lambda_{c}}, i=1,...,N.$
We first need to verify that $u=(u_i)=\phi^-$ is a lower solution of (\ref{eq211}) in the sense that
\begin{equation}\label{uppereq211-1}
d_iu''_i(\xi)-cu_i'(\xi)+f_i(u(\xi)) \geq 0, \;\; \xi \neq  \xi^*_i, i=1,...,N.
\end{equation}
It is clear that (\ref{uppereq211-1}) holds for $\xi > \xi^*_i$.
For  $\xi < \xi^*_i$,
$$u_i(\xi)=\nu^i_{\Lambda_{c}}e^{\Lambda_{c}\xi}-q\nu^i_{\gamma \Lambda_{c}}e^{\gamma\Lambda_{c}\xi}<\nu^i_{\Lambda_{c}}e^{\Lambda_{c}\xi}. $$
From Lemma \ref{estimageg}, we have, for  $\xi < \xi^*_i$ and $i=1,...,N$
\begin{equation}\label{uppereq40-1}
\begin{split}
& d_iu_i''(\xi)-cu_i(\xi)+f_i(u(\xi))\\
&\geq  d_i \Lambda_{c}^2\nu^i_{\Lambda_{c}}e^{\Lambda_{c}\xi}-c \Lambda_{c}\nu^i_{\Lambda_{c}}e^{\Lambda_{c}\xi}+\sum_{j=1}^n\partial_j f_i(0)\nu^j_{\Lambda_{c}}e^{\Lambda_{c}\xi}\\
& \quad -q\big(d_i (\gamma\Lambda_{c})^2\nu^i_{\gamma\Lambda_{c}}e^{\gamma\Lambda_{c}\xi}-c \gamma\Lambda_{c}\nu^i_{\gamma\Lambda_{c}}e^{\gamma\Lambda_{c}\xi}+\sum_{j=1}^n\partial_j f_i(0)\nu^j_{\gamma\Lambda_{c}}e^{\gamma\Lambda_{c}\xi}\big)\\
& \quad -\widehat{M}_i e^{2 \Lambda_{c}\xi}
\end{split}
\end{equation}
where $\widehat{M}_i=\sum_{j=1}^nb_{ij}(\nu^j_{\Lambda_{c}})^2>0.$
Note that $e^{(2-\gamma) \Lambda_{c}\xi}$ is bounded above for $\xi \leq \xi_i$. With the same argument in (\ref{uppereq40}), in particular, by Lemma \ref{lmeigen},
we get for sufficient large $q$ and $\xi \leq \xi_i$
\begin{equation}\label{uppereq40-2}
\begin{split}
& d_iu_i''(\xi)-cu_i(\xi)+f_i(u(\xi))\\
&\geq  q\big(c \gamma\Lambda_{c}- \gamma \Lambda_{c}\Phi(\gamma \Lambda_{c})\big)\nu^i_{\gamma\Lambda_{c}}e^{\gamma\Lambda_{c}\xi} -\widehat{M}_i e^{2 \Lambda_{c}\xi}\\
&=  q\big(c -\Phi(\gamma \Lambda_{c})\big)\gamma\Lambda_{c} \nu^i_{\gamma\Lambda_{c}}e^{\gamma\Lambda_{c}\xi} -\widehat{M}_i e^{2 \Lambda_{c}\xi}\\
&\geq  \Big(q\big(c -\Phi(\gamma \Lambda_{c})\big)\gamma\Lambda_{c} \nu^i_{\gamma\Lambda_{c}} -\widehat{M}_i e^{(2-\gamma) \Lambda_{c}\xi}\Big)e^{\gamma\Lambda_{c}\xi}\\
& \geq 0.
\end{split}
\end{equation}
Since $u'_i(\xi^*-) \leq u'_i(\xi^*+)=0$, the proof of Lemma 2.6 in Ma \cite{ma2001} immediately implies that
$$
\mathcal{T}_i[u](\xi) \geq u_i(\xi), \;\; \text{for all } \xi \in \mathbb{R}, i=1,...,N.
$$
Thus $u=(u_i)=\phi^+$ is a lower solution
of  (\ref{eq3fixpoint}). Note that the upper solutions in this paper are defined for equivalent integral equations other than differential equations  as in \cite{ma2001}.
\epf.

\subsection{Proof of Theorem \ref{th30}  (iii) when (\ref{eq1}) is cooperative }\label{proofMono}

In this section, we assume that (\ref{eq1}) is cooperative and prove Theorem \ref{th30} (iii). As we note in Remark \ref{rem2}, $f^{\pm}=f$.
Many results in this section are standard. See, for example, Ma \cite{ma2001,ma2007} and Wang \cite{Hwang2009,HwangSystemPDE}.
Define the following Banach space
$$
\mathcal{E}_{\varrho}= \{u=(u_i): u_i \in C(\mathbb{R}),\sup_{ \xi\in \mathbb{R}} \abs{u_i(\xi)}e^{-\varrho \xi} < \infty, i=1,...,N\}
$$
equipped with weighted norm
$$
\norm{u}_{\varrho}=\sum_{i=1}^N\sup_{ \xi\in \mathbb{R}} \abs{u_i(\xi)}e^{-\varrho \xi},
$$
where $C(\mathbb{R})$ is the set of all continuous functions on $\mathbb{R}$ and $\varrho$ is a positive constant such that $\varrho<\Lambda_{c}.$  It follows that
$\phi^+ \in \mathcal{E}_{\varrho}$ and $\phi^-\in \mathcal{E}_{\varrho}.$
Consider the following set
$$
\mathcal{A}=\{ u=(u_i): u_i \in C(\mathbb{R}) \in \mathcal{E}_{\varrho}, \phi_i^-(\xi) \leq u_i \leq \phi_i^+(\xi), \xi \in \mathbb{R}, i=1,...,N.\}
$$
We shall state the following lemmas. It is standard procedures to prove that $\mathcal{T}[u]$ is monotone, continuous and maps a bounded set in $\mathcal{A}$ into a compact set.
We omit their proofs here. The proof of compactness can be carried out two steps. First we show that $\mathcal{T}$ is  equicontinuous,  and then Ascoli's theorem and standard diagonal process
can be used to prove that  $\mathcal{T}$ is relatively compact  $\mathcal{E}_{\varrho}.$
The proofs of Lemmas \ref{monotone}, \ref{conunity}, \ref{compact} are almost identical
to those in  \cite{Hwang2009,HwangSystemPDE}.

\slm\label{monotone} Assume $(H1)-(H2)$ hold and $\partial_if_j\geq 0, i\neq j$ on $[0,k]$.  Then $\mathcal{T}$ defined in (\ref{eq3}) is monotone
and therefore $\mathcal{T}(\mathcal{A}) \subseteq \mathcal{A}$. Furthermore, $\mathcal{T}_i[u]$ is nondecreasing if $u \in \mathcal{A}$
and all of $u_i$ are nondecreasing.
\elm

\slm\mlabel{conunity} Assume $(H1)-(H2)$ hold. Then
$\mathcal{T}: \mathcal{A} \rightarrow \mathcal{E}_{\varrho}$ is continuous with the weighted norm $\norm{.}_{\varrho}$.
\elm
\slm\mlabel{compact} Assume $(H1)-(H2)$ hold. Then the set $\mathcal{T}(\mathcal{A})$ is relatively compact in $\mathcal{E}_{\varrho}.$
\elm

Now we are in a position to prove Theorem \ref{th30} when (\ref{eq1}) is cooperative.  Define the following iteration
\begin{equation}\label{iteration1}
u^1=(u_i^1)=\mathcal{T}[\phi^+], \;\; u_{n+1}=(u_i^n)=\mathcal{T}[u^n], n>1.
\end{equation}
From Lemmas \ref{upper}, \ref{sub}, \ref{monotone}, $u_n$ is nondecreasing on $\mathbb{R}$ and $$
\phi_i^-(\xi) \leq u_i^{n+1}(\xi) \leq u_i^n(\xi) \leq \phi_i^+(\xi), \xi \in \mathbb{R}, \;n \geq 1,n=1,...,N.
$$
By Lemma \ref{compact} and monotonicity of ($u_n$), there is $u \in \mathcal{A}$ such that $\lim_{n\to \infty}\norm{u_n-u}_{\varrho}=0$.  Lemma \ref{conunity}
implies that $\mathcal{T}[u]=u$. Furthermore, $u$ is nondecreasing. It is clear that $\lim_{\xi \to -\infty}u_i(\xi)=0,i=1,...,N$.
Assume that $\lim_{\xi \to \infty}u_i(\xi)=k_i',i=1,...,N$
$k'_i>0, i=1,...,N$ because of $u \in \mathcal{A}$. Applying the Dominated convergence theorem  to (\ref{eq3}), we get $k_i'=\frac{1}{\beta}(\beta k_i'+f_i(k_1',...,k'_n)$
By (H1), $k_i'=k_i$. Finally, note that
$$
\nu^i_{\Lambda_{c}}(e^{\Lambda_{c} \xi}-q\e^{\gamma \Lambda_{c}\xi}) \leq u_i(\xi) \leq \nu^i_{\Lambda_{c}}e^{\Lambda_{c}\xi}, \xi \in \mathbb{R}.
$$
We immediately obtain
\begin{equation}\label{aystom1}
\lim_{\xi \to -\infty}u_i(\xi)e^{-\Lambda_{c}\xi}=\nu^i_{\Lambda_{c}}, i=1,...,N.
\end{equation}
This completes the proof of Theorem \ref{th30} (iii) when (\ref{eq1}) is cooperative.

\subsection{Proof of Theorem \ref{th30} (iii)}\label{proofofTh1iii}
\pf
Theorem \ref{th30} (iii) is proved when (\ref{eq1}) is cooperative in the last section.  
Now we need to prove it in the general case(\ref{eq1}) is not necessarily cooperative. 
In order to find traveling waves for (\ref{eq1}), we will apply the Schauder's fixed point theorem.
Similar arguments can be found in the references, e.g.  \cite{ma2007} and \cite{HwangSystemPDE}.

Let $u=(u_i) \in \mathcal{A}$ and define two integral operators
$$\mathcal{T}^{\pm}[u]=(\mathcal{T}_i^{\pm}[u])$$
 for $f^{-}$ and $f^{+}$
\begin{equation}\label{eq+}
\begin{split}
&\mathcal{T}_i^{\pm}[u](\xi)\\
&=\frac{1}{d_i(\lambda_{1i}+\lambda_{2i})}[\int_{-\infty}^{\xi}e^{-\lambda_{1i}(\xi-s)}H_{i}^{\pm}(u(s))ds +\int^{\infty}_{\xi}e^{\lambda_{2i}(\xi-s)}H_i^{\pm}(u(s))ds]
\end{split}
\end{equation}
and
$$H_i^{\pm}(u(s))= \beta u_i(s)+ f_i^{\pm}(u_(s)).$$
As in Section \ref{proofMono}, both $\mathcal{T}^+$ and $\mathcal{T}^-$ are monotone.  In view of Section \ref{proofMono} and the fact that
$f^-$ is nondecreasing, there exists a nondecreasing fixed point $u^-=(u_i^-)$ of $\mathcal{T}^{-}$ such that
$\mathcal{T}^-[u^-]=u^-$, $\lim_{\xi \to \infty}u_i^{-}(\xi)=k_i^-, i=1,...,N$, and $\lim_{\xi \to -\infty}u_i^{-}(\xi)=0, i=1,...,N$. Furthermore,
$\lim_{\xi \to -\infty}u_i^{-}(\xi)e^{-\Lambda_{c}\xi}=\nu^i_{\Lambda_{c}}, i=1,...,N.$ According to Lemma \ref{upper},
$\phi^+$ (with $k$ being replaced
by $k^{\pm}$) is also a upper solution of $\mathcal{T}^{\pm}$
because the proof of Lemma \ref{upper} is still valid if $f$ is replaced by $f^{\pm}$.  Let
$$
\widetilde{\phi^+}(\xi)=(\widetilde{\phi^+_i}(\xi)),
$$
where
$$
\widetilde{\phi^+_i}(\xi)=\min\{k_i^+, \nu^i_{\Lambda_c}e^{\Lambda_{c}\xi}\}, i=1,...,N,\; \xi \in \mathbb{R};
$$
It follows that $u_i^-(\xi) \leq \widetilde{\phi^+_i}, \xi \in \mathbb{R}, i=1,...,N.$
Now let
\begin{equation}\label{definitionofB}
\mathcal{B}=\{ u: u=(u_i) \in \mathcal{E}_{\varrho}, u_i^-(\xi) \leq u_i(\xi) \leq \widetilde{\phi^+_i}(\xi), \xi \in ( -\infty, \infty), i=1,...,N\},
\end{equation}
where $\mathcal{E}_{\varrho}$ is defined in Section \ref{proofMono}.
It is clear that $\mathcal{B}$ is a bounded nonempty closed convex subset in $\mathcal{E}_{\varrho}$.
Furthermore, we have, for any $u=(u_i) \in \mathcal{B}$
$$
u_i^- = \mathcal{T}_i^-[u^-]\leq \mathcal{T}_i^-[u] \leq \mathcal{T}_i[u] \leq \mathcal{T}_i^+[u]\leq \mathcal{T}_i^+[\widetilde{\phi^+}] \leq \widetilde{\phi^+_i}, i=1,...,N.
$$
Therefore,  $\mathcal{T}: \mathcal{B} \rightarrow \mathcal{B}$. Note that the proofs of Lemmas \ref{conunity}, \ref{compact} are valid if (\ref{eq1}) is
not cooperative. In the same way as in Lemmas \ref{conunity}, \ref{compact} , we can show that
$\mathcal{T}: \mathcal{B} \rightarrow \mathcal{B}$ is continuous and maps bounded sets into compact sets.
Therefore, the Schauder Fixed
Point Theorem shows that the operator $\mathcal{T}$ has a fixed point $u$ in $\mathcal{B}$, which is a traveling wave
solution of (\ref{eq1}) for $c>c^*$. Since $u_i^-(\xi) \leq u_i(\xi) \leq \widetilde{\phi^+_i}(\xi), \xi \in ( -\infty, \infty), i=1,...,N$,
it is easy to see that for $i=1,...,N$,
$\lim_{\xi \to -\infty}u_i(\xi)=0$, $\lim_{\xi \to -\infty}u_i(\xi)e^{-\Lambda_{c}\xi}=\nu^i_{\Lambda_{c}}$,
$$k^-\leq \liminf_{\xi \to \infty}u(\xi) \leq \limsup_{\xi \to \infty}u(\xi)\leq k^+$$ and
$0 < u_i^-(\xi) \leq  u_i(\xi) \leq k_i^+, \xi \in ( -\infty, \infty)$.
\epf

\subsection{Proof of Theorem \ref{th30} (iv)} \label{proofofTh1iv}
\pf We adopt the limiting approach in \cite{BrownCarr1977} to prove  Theorem \ref{th30} (iv).  For each $n \in \mathbb{N}$,
choose $c_n >c^*$ such that $\lim_{n\to \infty} c_n=c^*.$ According to Theorem \ref{th30} (iii), for each $c_n$
there is a traveling wave solution $u_n=(u^n_i)$ of (\ref{eq1}) such that
$$
u_n=\mathcal{T}[u_n](\xi).
$$
and
$$k^-_i\leq \liminf_{\xi \to \infty}u^n_i(\xi) \leq \limsup_{\xi \to \infty}u^n_i(\xi)\leq k_i^+, i=1,...,N.$$
It can be shown that $(u_n)$ is equicontinuous and uniformly bounded on $\mathbb{R}$ (see, e.g. \cite[Lemma 5.3]{Hwang2009}), the Ascoli's theorem implies that
there is vector valued continuous function $u=(u^i)$ on $\mathbb{R}$ and subsequence $(u_{n_m})$ of $(u_n)$ such that $$
\lim_{m \to \infty} u_{n_m}(\xi)=u(\xi)
$$
uniformly in $\xi$ on any compact interval of $\mathbb{R}$. Further in view of the dominated convergence theorem we have
$$
u=\mathcal{T}[u](\xi).
$$
Here the underlying $\lambda_{1i}, \lambda_{2i}$ of $\mathcal{T}$ is dependent on $c$ and continuous functions of $c$.
Thus $u$ is a traveling solution of  (\ref{eq1}) for $c=c^*$. Since, for each $c_n$, $u_n \in \mathcal{B}$ where $\mathcal{B}$ is defined
in (\ref{definitionofB}),  it is easy to see that $u$ satisfies
$$k^-\leq \liminf_{\xi \to \infty}u(\xi) \leq \limsup_{\xi \to \infty}u(\xi)\leq k^+$$
Because of the translation invariance of $u_n$, we always can assume that $u_n(0) \leq \frac{1}{2}k^-$ for all $n$.
Consequently $u$ is not a constant traveling solution of (\ref{eq1}).
\epf

\subsection{Proof of Theorem \ref{th30} (v)} \label{proofofTh1v}
\pf
 Suppose, by contradiction, that for some $c \in (0, c^*)$, (\ref{eq1})
has a traveling wave $u(x,t)=u(x + ct)$  with $ \liminf_{\xi \to \infty} u(\xi)>>0$ and $u(-\infty)=0.$  Thus $u(x,t)=u(x + ct)$ can be larger than
a positive vector with arbitrary length. It follows from Theorem \ref{th30} (ii)
$$
\liminf_{t \to \infty} \inf_{\abs{x}  \leq ct} u(x, t) \geq k^->>0, \text{ for } 0< c< c^*
$$
Let $\hat{c} \in (c,c^*)$ and $x=\hat{c}t.$ Then $$
\lim_{t\to \infty} u\big(-(\hat{c}-c)t\big) =\lim_{t\to \infty} u(-\hat{c}t, t) \geq \liminf_{t \to \infty} \inf_{\abs{x}  \leq t\hat{c}} u(x,t) >>0.
$$
However,
$$
\lim_{t \to \infty}u\big(-(\hat{c}-c)t\big)=u(-\infty)=0,
$$
which is a contradiction.
\epf

\end{document}